\documentclass[12pt,thmsa]{article}
\def \eq {\begin{equation}}
\def \fim-eq {\end{equation}}
\begin{document}

\author{E. S. Guerra \\
%EndAName
Departamento de F\'{\i}sica \\
Universidade Federal Rural do Rio de Janeiro \\
Cx. Postal 23851, 23890-000 Serop\'edica, RJ, Brazil \\
email: emerson@ufrrj.br\\
}
\title{ON THE COMPLEMENTARITY PRINCIPLE AND THE UNCERTAINTY PRINCIPLE}
\maketitle

\begin{abstract}
\noindent\ We present a scheme in which we investigate the two-slit
experiment and we show that the principle of complementarity is more
fundamental then the uncertainty principle.\ \newline

\end{abstract}

The investigation of the double slit experiment in which we have two
micromaser cavities, each one associated with one of the slits, was proposed
before by Scully and Walther in a series of very interesting articles where
they investigate the principle of complementarity and the uncertainty
principle \cite{DSSW, Baggott} and they conclude that the principle of
complementarity is more fundamental than the uncertainty principle. \ The
same conclusion can be reached very easily using the scheme presented below.
\ We are going to consider a screen with two slits $SL1$ (at $\zeta _{1}$)
and $SL2$ \ (at $\zeta _{2}$) with two cavities $C1$ and $C2$ behind
respectively each slit and prepared respectively in an even coherent state $%
|+\rangle _{1}$ and an odd coherent state $|-\rangle _{2}$, where 
\begin{equation}
\mid \pm \rangle _{k}=\mid \alpha _{k}\rangle \pm \mid -\alpha _{k}\rangle
\label{EOCS}
\end{equation}%
\cite{EvenOddCS}, through which fly Rydberg atoms of relatively long
radiative lifetimes \cite{Rydat}. We also assume perfect \ microwave
cavities, that is, we neglect effects due to decoherence. Concerning this
point, it is worth to mention that nowadays it is possible to build up
niobium superconducting cavities with high quality factors $Q$. It is
possible to construct cavities with quality factors $Q\sim 10^{8}$ \cite%
{haroche}. Even cavities with quality factors as high as $Q\sim 10^{12}$
have been reported \cite{walther}, which, for frequencies $\nu \sim 50$ GHz
gives us a cavity field lifetime of the order of a few seconds.

Let us first show \ a possible way of preparing the cavities in the states (%
\ref{EOCS}). Suppose we prepare cavity $Ck$ initially in a coherent state $%
|i\alpha _{k}\rangle $. Then, we prepare a two-level atom $A0$ in a coherent
superposition, preparing it initially in state $\mid f_{0}\rangle ,$ passing
it through a first Ramsey zone $R1$ where%
\begin{equation}
R_{1}=\frac{1}{\sqrt{2}}\left[ 
\begin{array}{cc}
c_{f} & c_{e} \\ 
-c_{e} & c_{f}%
\end{array}%
\right] ,
\end{equation}%
and we get%
\begin{equation}
\mid \psi \rangle _{0}=c_{e}\mid e_{0}\rangle +c_{f}\mid f_{0}\rangle .
\end{equation}%
After that, $A0$ flies through cavity $Ck$. The $\mid e_{0}\rangle
\rightleftharpoons \mid f_{0}\rangle $ transition is far from resonance with
the cavity central frequency such that only virtual transitions occur
between these states and the interaction of the atom with the cavity mode in 
$C1$ is described by the time evolution operator \cite{Orszag}. 
\begin{equation}
U(t)=e^{-i\varphi (a^{\dagger }a+1)}\mid e\rangle \langle e\mid +e^{i\varphi
a^{\dagger }a}\mid f\rangle \langle f\mid ,  \label{U1}
\end{equation}%
where $a$ $(a^{\dagger })$ is the annihilation (creation) operator, $\varphi
=g^{2}\tau /$ $\Delta $, \ $g$ is the coupling constant, $\Delta =\omega
_{e}-\omega _{f}-\omega $ is the detuning \ where \ $\omega _{e}$ and $%
\omega _{f}$ \ are the frequency of the upper and lower levels respectively
and $\omega $ is the cavity field frequancy and $\tau $ is the atom-field
interaction time. After the atom passes through $C1$ the state of the system 
$A0-C1$, for $\varphi =\pi /2$, is given by 
\[
\mid \psi \rangle _{A0-C1}=-ic_{e}\mid e_{0}\rangle \mid \alpha _{k}\rangle
+c_{f}\mid f_{0}\rangle \mid -\alpha _{k}\rangle _{.} 
\]%
Then, the atom enters a second Ramsey zone $R2$ where we have 
\begin{equation}
R_{2}=\frac{1}{\sqrt{2}}\left[ 
\begin{array}{cc}
1 & -i \\ 
i & 1%
\end{array}%
\right] ,
\end{equation}%
that is,%
\begin{eqnarray}
&\mid &e_{0}\rangle \rightarrow \frac{1}{\sqrt{2}}(\mid e_{0}\rangle +i\mid
f_{0}\rangle ),  \nonumber \\
&\mid &f_{0}\rangle \rightarrow \frac{1}{\sqrt{2}}(-i\mid e_{0}\rangle +\mid
f_{0}\rangle ).
\end{eqnarray}%
After the atom cross the Ramsey zone $R2$, the state of the system $A0+Ck$
is given by 
\begin{eqnarray*}
|\psi \rangle _{A0-Ck} &=&\frac{1}{\sqrt{2}}\big[-ic_{e}\mid \alpha
_{k}\rangle -ic_{f}\mid -\alpha _{k}\rangle \big]\mid e_{0}\rangle \\
&&+\frac{1}{\sqrt{2}}\big[c_{e}\mid \alpha _{k}\rangle +c_{f}\mid -\alpha
_{k}\rangle \big]\mid f_{0}\rangle ,
\end{eqnarray*}%
and for $c_{e}=c_{f}$ we get $|\psi \rangle _{Ck}=\mid +\rangle _{k}$ \ \ if
we detect $\mid e_{0}\rangle $ \ or $\mid f_{0}\rangle $ and for $%
c_{e}=-c_{f}$ we get $|\psi \rangle _{Ck}=\mid -\rangle _{k}$ \ \ if we
detect $\mid e_{0}\rangle $ \ or $\mid f_{0}\rangle $.

Now, for a three-level lambda atom interacting with the electromagnetic
field inside a cavity where the \ upper and the two degenerated lower states
are $|a\rangle ,$ $|b\rangle $ and $|c\rangle $ respectively, and for which
the $|a\rangle \rightleftharpoons |c\rangle $ and $|a\rangle
\rightleftharpoons |b\rangle $ transitions are in the far from resonance
interaction limit, the time evolution operator $U(t)$ for the atom-field
interaction in a cavity $Ck$ is given by \cite{Knight}%
\begin{eqnarray}
U(\tau ) &=&\frac{1}{2}(e^{i\varphi a_{k}^{\dagger }a_{k}}+1)|b\rangle
\langle b|+\frac{1}{2}(e^{i\varphi a_{k}^{\dagger }a_{k}}-1)|b\rangle
\langle c|\ +  \nonumber \\
&&\frac{1}{2}(e^{i\varphi a_{k}^{\dagger }a_{k}}-1)|c\rangle \langle b|+%
\frac{1}{2}(e^{i\varphi a_{k}^{\dagger }a_{k}}+1)|c\rangle \langle c|.
\end{eqnarray}%
where $a_{k}$ $(a_{k}^{\dagger })$ is the annihilation (creation) operator
for the field in cavity $Ck$, $\varphi =2g^{2}\tau /$ $\Delta $, \ $g$ is
the coupling constant, $\Delta =\omega _{a}-\omega _{b}-\omega =\omega
_{a}-\omega _{c}-\omega $ is the detuning where \ $\omega _{a}$, $\omega _{b}
$ and $\omega _{c}$\ are the frequency of the upper and \ of the two
degenerate lower levels respectively and $\omega $ is the cavity field
frequancy and $\tau $ is the atom-field interaction time. For $\varphi =\pi $%
, we get 
\begin{equation}
U(\tau )=-\exp \left( i\pi a_{k}^{\dagger }a_{k}\right) |a\rangle \langle
a|+\Pi _{k,+}|b\rangle \langle b|+\Pi _{k,-}|b\rangle \langle c|\ +\Pi
_{k,-}|c\rangle \langle b|+\Pi _{k,+}|c\rangle \langle c|,  \label{UlambdaPi}
\end{equation}%
where 
\begin{eqnarray}
\Pi _{k,+} &=&\frac{1}{2}(e^{i\pi a_{k}^{\dagger }a_{k}}+1),  \nonumber \\
\Pi _{k,-} &=&\frac{1}{2}(e^{i\pi a_{k}^{\dagger }a_{k}}-1),  \label{pi+-}
\end{eqnarray}%
and we have 
\begin{eqnarray}
\Pi _{k,+}|+\rangle _{k} &=&|+\rangle _{k},  \nonumber \\
\Pi _{k,+}|-\rangle _{k} &=&0,  \nonumber \\
\Pi _{k,-}|-\rangle _{k} &=&-|-\rangle _{k},  \nonumber \\
\Pi _{k,-}|+\rangle _{k} &=&0,
\end{eqnarray}%
which are easily obtained from Eqs. (\ref{pi+-}) and (\ref{EOCS}) using $%
e^{za_{k}^{\dagger }a_{k}}|\alpha _{k}\rangle =|e^{z}\alpha _{k}\rangle $ 
\cite{Louisell}. Let us assume that we send three-level lambda atoms through
the slits and cavities. Consider an atom $A1$ prepared in the state $%
|b_{1}\rangle $ flying through the double slit. Before $A1$ crosses the
cavities we have%
\begin{equation}
|\psi \rangle _{A1-SL1-SL2}=\frac{1}{\sqrt{2}}(|\zeta _{1}\rangle +|\zeta
_{2}\rangle )|+\rangle _{1}|-\rangle _{2}\mid b_{1}\rangle
\end{equation}%
and after it has interacted with $C1$ and $C2,$ taking into account (\ref%
{UlambdaPi}),%
\begin{equation}
|\psi (t_{0})\rangle _{A1-C1-C2}=\frac{1}{\sqrt{2}}(|\zeta _{1}\rangle \mid
b_{1}\rangle -|\zeta _{2}\rangle \mid c_{1}\rangle )|+\rangle _{1}|-\rangle
_{2}.
\end{equation}%
Now, writing \ $\psi (x,t)_{A1-C1-C2}=\langle x|U(t,t_{0})|\psi
(t_{0})\rangle _{A1-C1-C2}$, $\psi _{1}(x,t)=\langle x|U(t,t_{0})|\zeta
_{1}\rangle $ and $\psi _{2}(x,t)=\langle x|U(t,t_{0})|\zeta _{2}\rangle ,$
where $|x\rangle $ is a point on a screen in front of the double slit screen
at a certain distance $L$ from it, we have%
\begin{equation}
\ \psi (x,t)_{A1}=\frac{1}{\sqrt{2}}\{\psi _{1}(x,t)\mid b_{1}\rangle -\psi
_{2}(x,t)\mid c_{1}\rangle \}.
\end{equation}%
and%
\begin{equation}
\ \left\vert \psi (x,t)_{A1}\right\vert ^{2}=\frac{1}{2}[\left\vert \psi
_{1}(x,t)\right\vert ^{2}+\left\vert \psi _{2}(x,t)\right\vert ^{2}]
\end{equation}%
since $\langle c_{1}\mid b_{1}\rangle =0$ and if there were no cavities we
would obtain%
\begin{equation}
\ \left\vert \psi (x,t)_{A1}\right\vert ^{2}=\frac{1}{2}[\left\vert \psi
_{1}(x,t)\right\vert ^{2}+\left\vert \psi _{2}(x,t)\right\vert ^{2}+2{Re}\{\psi _{1}^{\ast }(x,t)\psi _{2}(x,t)\}]
\end{equation}%
which presents interference fringes. Therefore, when we place cavities $C1$
and $C2$ prepared in the states $|+\rangle _{1}$ and $|-\rangle _{2}$
respectively, the interference fringes are washed out. \ This happens
because the parity information of the cavities is transferred to the
internal state of the atom. Notice that if we detect the atomic state of $A1$
after it has crossed the slits and before it strikes the detection screen at 
$|x\rangle $ and we find $\mid b_{1}\rangle ,$ we can say that the atom has
passed through slit $SL1,$ and if we detect $\mid c_{1}\rangle ,$ we can say
that the atom has passed through slit $SL2$ and we get which-path
information (particle behavior) detecting the atomic state. That is,
assuming that the detection of the internal states does not disturb the
external state of motion of the centre of mass of the atom, in the case we
detect $\mid b_{1}\rangle $ \ we get 
\begin{equation}
\left\vert \psi (x,t)_{A1}\right\vert ^{2}=\left\vert \psi
_{1}(x,t)\right\vert ^{2},
\end{equation}%
and in the case we detect $\mid c_{1}\rangle $ \ we get%
\begin{equation}
\left\vert \psi (x,t)_{A1}\right\vert ^{2}=\left\vert \psi
_{2}(x,t)\right\vert ^{2}.
\end{equation}%
Therefore, the cavities allow us to get which-path information. Notice that
as the atom-field interaction is dispersive, we can state that the
uncertainty principle plays no role at all. Therefore, we get the same
conclusion of \ Scully and Walther \cite{DSSW}, that is, the complementarity
principle is more fundamental than the uncertainty principle since as the
experiment cited and the present experiment represent a way around the
uncertainty principle.

Now, assume that we place a Ramsey cavity on the way of the atom just before
it passes through the slits and cavities so that the atomic state before the
atom crosses the cavities be 
\begin{equation}
|\psi \rangle _{A1}=\frac{1}{\sqrt{2}}(\mid b_{1}\rangle +\mid c_{1}\rangle
).
\end{equation}%
In this case the initial state of the system is 
\begin{equation}
|\psi \rangle _{A1-SL1-SL2}=\frac{1}{2}(|\zeta _{1}\rangle +|\zeta
_{2}\rangle )|+\rangle _{1}|-\rangle _{2}(\mid b_{1}\rangle +\mid
c_{1}\rangle )
\end{equation}%
and after the atom crosses the cavities%
\begin{equation}
|\psi \rangle _{A1-C1-C2}=\frac{1}{2}[|\zeta _{1}\rangle (\mid b_{1}\rangle
+\mid c_{1}\rangle )-|\zeta _{2}\rangle (\mid c_{1}\rangle +\mid
b_{1}\rangle )]|+\rangle _{1}|-\rangle _{2}
\end{equation}%
and therefore we have%
\begin{equation}
\ \left\vert \psi (x,t)_{A1}\right\vert ^{2}=\frac{1}{4}\{\left\vert \psi
_{1}(x,t)\right\vert ^{2}+\left\vert \psi _{2}(x,t)\right\vert ^{2}-2{Re%
}\{\psi _{1}^{\ast }(x,t)\psi _{2}(x,t)\}]
\end{equation}%
and we have interference as we would expect since the atom enters the
cavities in an atomic state superposition of their internal states and no
information about the state of the cavities can be transferred to the atom
which exits the cavities also in an atomic state superposition which do not
allows us to get which-path information detecting the atomic state. Note
that, in the case we do not have a Ramsey cavity and we prepare the atom,
for instance, in state $\mid b_{1}\rangle ,$ we know that it is in this
state and, if we accept the theoretical formalism of quantum mechanics as
correct, according to it we know that in this case, as we have seen above,
the parity information of the cavities is transferred to the atom allowing
us to get which-path information.

\ Let us consider now a scheme of atomic states similar the one used above
to prepare the cavities in an even or an odd coherent state, but now we take
a three-level cascade atom \ with $\mid e\rangle ,\mid f\rangle $ and $\mid
g\rangle $ being the upper, intermediate and lower atomic states. Again we
assume that the transition $\mid f\rangle \rightleftharpoons \mid e\rangle $
is far enough from resonance with the cavity central frequency such that
only virtual transitions occur between these states. In addition we assume
that the transition $\mid e\rangle \rightleftharpoons \mid g\rangle $ is
highly detuned from the cavity frequency so that there will be no coupling
with the cavity field (only the states $\mid f\rangle $ and $\mid e\rangle $
interact with the field in the cavity). Here we are going to consider the effect of
the atom-field interaction taking into account only levels $\mid f\rangle $
and $\mid g\rangle $. We do not consider level $\mid e\rangle $ since it
will not play any role in our scheme. Therefore, we have effectively a
two-level system involving states $\mid f\rangle $ and $|g\rangle $.
Considering levels $\mid f\rangle $ and $\mid g\rangle $ and taking into
account (\ref{U1}), we can write an effective time evolution operator 
\begin{equation}
U_{k}(t)=e^{i\varphi a^{\dagger }a}\mid f\rangle \langle f\mid +|g\rangle
\langle g\mid ,  \label{U2}
\end{equation}%
where the second term above was put by hand just in order to take into
account the effect of level $\mid g\rangle $. Let us assume that atom $A1$
is prepared in state $\mid g\rangle $ and, on the way to the screen with
two\ slits and two cavities, there is a Ramsey cavity $R1$ were the atomic
states are rotated according to%
\begin{equation}
R_{1}=\frac{1}{\sqrt{2}}\left[ 
\begin{array}{cc}
1 & 1 \\ 
-1 & 1%
\end{array}%
\right] ,
\end{equation}%
that is, after $A1$ crosses this Ransey cavity we have%
\begin{equation}
\mid \psi \rangle _{A1}=\frac{1}{\sqrt{2}}(\mid f_{1}\rangle +\mid
g_{1}\rangle ),
\end{equation}%
and, for $\varphi =\pi $, after $A1$ flies through slit $SL1$ and cavity $C1$
and through slit $SL2$ and cavity $C2$, we have%
\begin{equation}
\mid \psi (t_{0})\rangle _{A1-C1-C2}=\frac{1}{2}\{|\zeta _{1}\rangle (\mid
f_{1}\rangle +\mid g_{1}\rangle )+|\zeta _{2}\rangle (-\mid f_{1}\rangle
+\mid g_{1}\rangle )\}|+\rangle _{1}|-\rangle _{2},
\end{equation}%
and taking into account the time evolution operator $U(t,t_{0})$%
\begin{equation}
\left\vert \psi _{1}(x,t)_{A1}\right\vert ^{2}=\frac{1}{4}[\left\vert \psi
_{1}(x,t)\right\vert ^{2}+\left\vert \psi _{2}(x,t)\right\vert ^{2}],
\end{equation}%
and we have no interference fringes since $(-\langle f_{1}|+\langle g_{1}|)$ 
$(\mid f_{1}\rangle +\mid g_{1}\rangle )=0$. Now, if we do not have $R1$ on
the way of the atom to the two-slit screen we get%
\begin{equation}
\mid \psi (t_{0})\rangle _{A1-C1-C2}=\frac{1}{2}\{|\zeta _{1}\rangle \mid
g_{1}\rangle -|\zeta _{2}\rangle \mid g_{1}\rangle \}|+\rangle _{1}|-\rangle
_{2}
\end{equation}%
and therefore, 
\begin{equation}
\ \left\vert \psi (x,t)_{A1}\right\vert ^{2}=\frac{1}{2}[\left\vert \psi
_{1}(x,t)\right\vert ^{2}+\left\vert \psi _{2}(x,t)\right\vert ^{2}-2{Re%
}\{\psi _{1}^{\ast }(x,t)\psi _{2}(x,t)\}],
\end{equation}%
and we have interference fringes. In the case we have \ the Ramsey cavity on
the way of the atom to the two-slit screen the atom enters the cavities $C1$
and $C2$ in an atomic state superposition and now the cavities transfer
their parity information to the atom. We can see that we can get which-path
information in this case if we consider another Ramsey cavity $R2$ where 
\begin{equation}
R_{2}=\frac{1}{\sqrt{2}}\left[ 
\begin{array}{cc}
1 & 1 \\ 
-1 & 1%
\end{array}%
\right] ,
\end{equation}%
and%
\begin{eqnarray}
\frac{1}{\sqrt{2}}( &\mid &f_{1}\rangle +\mid g_{1}\rangle )\longrightarrow
\mid f_{1}\rangle , \\
\frac{1}{\sqrt{2}}(- &\mid &f_{1}\rangle +\mid g_{1}\rangle )\longrightarrow
\mid g_{1}\rangle 
\end{eqnarray}%
just behind $C1$ or $C2$. If $R2$ is just after cavity $C1$ we can detect
atom $A1$ in state $\mid f_{1}\rangle $ and we know that the atom has passed
through slit $SL1$ and if $R2$ is just after cavity $C2$ we can detect atom $%
A1$ in state $\mid g_{1}\rangle $ and we know that the atom has passed
through slit $SL2$, that is, we get which-path information. In the case we
do not have $R1$ on the way of the atom to the slits, it is clear that we
cannot get which-path information.

We should point out that the field of the cavities is not disturbed in any
way \ in the above schemes and therefore, these experiments could be
performed with any number of atoms which would allow us to get
experimentally a two peaks pattern (particle behavior) or a wiggly pattern
(wave behavior). Therefore, aside technical difficulties, these experiments
could not be considered \textit{gedanken} experiments. A technical problem
in these experiments is related to the fact that the separation between the
slits should be very small and one should work also with very small
cavities. Although this is not an easy problem to deal with, we think that
the schemes we have discussed are at least of academic interest. We should
point out also that many experiments which were considered as \textit{%
gedanken }experiments in the past turned out to be realized in laboratory
nowadays. \ 

We intend to publish a further investigation along this line elsewhere.

\end{document}